\documentstyle[epsf]{mn}
\title[$\eta$ Carinae]
{Variability of $\eta$ Carinae III}
\author[M.W. Feast et al.]
{ Michael Feast$^{1}$, Patricia Whitelock$^{2}$ and Freddy Marang$^{2}$\\
$^{1}$ Astronomy Department, University of Cape Town, Rondebosch
7701, South Africa.\\
email: mwf@artemisia.ast.uct.ac.za\\
$^{2}$ South African Astronomical Observatory, PO Box 9, 
Observatory, 7935, South Africa.\\
email: paw@saao.ac.za; fm@saao.ac.za\\} 

\begin{document}

\maketitle

\begin{abstract}
 Spectra (1951-78) of the central object in $\eta$ Car, taken by A.D.
Thackeray, reveal three previously unrecorded epochs of low excitation. 
Since 1948, at least, these states have occurred regularly in the 2020 day
cycle proposed by Damineli et al. They last about 10 percent of each cycle.
Early slit spectra (1899-1919) suggest that at that time the object was
always in a low state. $JHKL$ photometry is reported for the period
1994-2000. This shows that the secular increase in brightness found in
1972-94 has continued and its rate has increased at the shorter wavelengths.
Modulation of the infrared brightness in a period near 2020 days continues.
There is a dip in the $JHKL$ light curves near 1998.0, coincident with a dip
in the X-ray light curve. Evidence is given that this dip in the infrared
repeats in the 2020 day cycle. As suggested by Whitelock \& Laney, the dip
is best interpreted as an eclipse phenomenon in an interacting binary
system; the object eclipsed being a bright region (`hot spot'), possibly on
a circumstellar disc or produced by interacting stellar winds. The eclipse
coincides in phase and duration with the state of low excitation.  It is
presumably caused by a plasma column and/or by one of the stars in the
system.
\end{abstract}


\section{Introduction}
 Despite a very extensive literature there is as yet no general consensus on
the nature of $\eta$ Carinae and the cause of the Great Eruption (c.
1838 - 1858) (see Morse et al. 1999 
and Davidson \& Humphreys 1997 for recent summaries). The present paper
is the third in a series reporting observations of this object. The earlier
papers are Whitelock et al. 1983 (henceforth Paper~I) and Whitelock et al.
1994 (henceforth Paper~II). Paper~I reported and discussed infrared, $JHKL$,
photometry of the object in the years 1972 to 1982 as well as optical and
infrared spectroscopy. Paper~II continued the photometric series up to 1994
and showed that, besides a gradual increase in brightness over the 
entire time interval 1972 to 1994, there was evidence for variations on a
time scale of about 5 years. 
This was probably the first convincing evidence for periodicity or
quasi-periodicity in the system.
Damineli (1996) combined these data with
observations of the variations in the He\,{\sc i} 10830A emission line and
other spectroscopic results in the literature to suggest that there was a
rather stable periodicity of 5.52 years. Later Damineli et al. (1997) found
evidence for a variation in the radial velocity of Pa$\gamma$ emission in
this same period. They interpreted this as showing that $\eta$ Car was a
spectroscopic binary in a highly eccentric orbit and they suggested that
various phenomena were associated with periastron passage (see also Damineli
et al. 2000). However, Davidson et al. (2000) report HST observations of the
central object, which is not resolved from the ground. These do not confirm
the velocity behaviour predicted by the Damineli et al. orbit. In the
present paper we report an examination of spectroscopic material,
particularly that collected by A.D. Thackeray over the years 1951 to 1978,
and also give the results of the continued monitoring of $\eta$ Car in
$JHKL$. We discuss these data, as well as published material, with
particular reference to the light they throw on the periodicity and nature
of the object.

\section{Periodic Spectroscopic Variations}
 Gaviola (1953) found that there was a marked difference between a spectrum
of $\eta$ Car taken in 1948 and those taken in 1947 and 1949. He\,{\sc i},
[Ne\,{\sc iii}] and various other emission lines of above average excitation
were missing in 1948. A similar low excitation event was found to occur in
1965 (Thackeray 1967, Rodgers \& Searle 1967). Paper~I reported that the
object was again in a low excitation state in 1981 at a time when $JHKL$
were near maximum brightness in the $\sim$ 5 year cycle. In the present
section we discuss all the spectroscopic studies since 1947 relevant to the
high/low excitation state which are known to us, including three previously
unreported low states seen on plates taken by A.D. Thackeray.

Thackeray obtained a large number of spectra both of the `central object'
(as seen from the ground) and portions of the surrounding (bipolar)
`homunculus' and other nebulosity.  He published discussions of only a small
part of this material though it seems likely that had he lived he would have
consolidated his work on this object as he did in the case of RR Tel
(Thackeray 1977). There are two main series of his (photographic) spectra.
One of these was with the two-prism Cassegrain spectrograph on the Radcliffe
(now SAAO) 1.9m reflector. This began on 1951 April 12, soon after the
spectrograph was installed and continued, with intermissions, till 1978
February 15, a few days before Thackeray died. The other series, with the
coud\'{e} grating spectrograph on the same telescope extended from 1960 May
16 to 1974 January 8. All these plates are now in the archives of the SAAO
and have been examined. Table 1 lists only those spectra of the central
object which are suitable for determining whether it was in a high or low
excitation state.

It was found that the spectra could be readily classified using a hand lens
and/or a Hartmann spectrocomparator. In this classification one is concerned
with the sharp emission lines, not the broad emission lines which underlie
many of them. The excitation state was judged using some or all of the
following line ratios; 
4471He\,{\sc i}/4475[Fe\,{\sc ii}]; 
4658[Fe\,{\sc III}]/4667[Fe\,{\sc ii}]; 
5876He\,{\sc I}/5894Na\,{\sc i}; 
5754[N\,{\sc ii}]/5748[Fe\,{\sc ii}]; 
6312[S\,{\sc iii}]/6318Fe\,{\sc ii}; 
and the presence or absence of 3869,3967[Ne\,{\sc iii}]. The strength of the Balmer absorption lines was
also taken into account. Damineli's (1996) work indicates that the
equivalent width of 10830He\,{\sc i} varies continuously through the 5.52
year cycle as does the $JHKL$ brightness. It may be somewhat surprising
therefore that it was possible on the basis of the above criteria to
classify the sharp-line spectra unambiguously as belonging to either the
normal (high excitation) state or the low state. However, the 10830He\,{\sc
i} does not necessarily arise entirely in the same region as the sharp line
spectra. 
Furthermore, as the referee has pointed out to us, the time interval
over which 10830A practically disappears is relatively short and well
defined.
There is only one spectrum (Cb1567) in which the ratio
4471He\,{\sc i}/4475[Fe\,{\sc ii}] suggests a possible intermediate state
and even this is somewhat uncertain. This is not to say that all the spectra
assigned to a particular state are identical. Differences between such
spectra are suspected and indeed were already noted between the high states
of 1947 and 1949 by Gaviola (1953). However, any such differences are minor
compared with the high/low state differences.

\begin{table}
\centering
\caption{Spectral State (Thackeray)}
\begin{tabular}{lrrl}
\multicolumn{1}{c}{Plate}&{Date}&{Phase}&{State}\\
& & & \\
Ca9     & 12.4.51  & 1.529 & H \\
Cc40    & 29.4.51  & 1.537 & H \\
Cc63    & 8.5.51   & 1.542 & H \\
Cb99    & 29.5.51  & 1.552 & H  \\
Cb100   & 29.5.51  & 1.552 & H \\
Ca109   & 2.6.51   & 1.554 & H \\
Cc135   & 11.6.51  & 1.559 & H \\
Ca142   & 14.6.51  & 1.560 & H \\
Cb204   & 7.7.51   & 1.572 & H \\
Cb208   & 10.7.51  & 1.573 & H \\
Ca896   & 26.6.52  & 1.747 & H \\
Cb1567  & 28.6.53  & 1.929 & I? \\ 
Cb1821  & 30.12.53 & 2.021 & L \\  
Cb1960  & 12.5.54  & 2.086 & L \\
Cc2313  & 24.2.55  & 2.229 & H \\
Ca2385  & 6.4.55   & 2.249 & H \\
Ca2430  & 11.5.55  & 2.267 & H \\
Cb2889  & 15.5.56  & 2.450 & H \\
Cb3246  & 14.5.59  & 2.991 & L \\
Cb4291  & 1.6.59   & 3.000 & L \\
Cc4661  & 6.2.60   & 3.124 & H \\
Cb4714  & 23.3.60  & 3.147 & H \\
Cb5086  & 21.5.61  & 3.357 & H \\
Cb5263  & 3.1.62   & 3.469 & H \\
Cb5787  & 2.5.63   & 3.709 & H \\
Cb6193  & 2.3.64   & 3.860 & H \\
Cc6713  & 24.3.65  & 4.051 & L \\
Cc7171  & 2.3.66   & 4.221 & H? \\
Cc7210  & 22.3.66  & 4.231 & H \\
Cb8205  & 4.6.68   & 4.629 & H \\
Cc8843a & 25.5.70  & 4.986 & L \\
Cb8846b & 4.6.70   & 4.991 & L \\
Cb9251  & 15.2.78  & 6.383 & H \\
 & & & \\
DZ10    & 16.5.60  & 3.174 & H \\
DZ31    & 18.5.60  & 3.175 & H \\
DZ43    & 8.6.60   & 3.185 & H \\
DZ52    & 9.6.60   & 3.185 & H \\
DY81    & 8.7.60   & 3.200 & H \\
DZ177   & 31.1.61  & 3.302 & H \\
DZ178   & 31.1.61  & 3.302 & H \\
DY646   & 18.5.62  & 3.536 & H \\
DZ658   & 20.5.62  & 3.537 & H \\
DZ1310  & 19.3.65  & 4.049 & L \\
DZ3041  & 2.1.72   & 5.276 & H \\
DZ3601  & 14.6.73  & 5.538 & H \\
DY3726  & 8.1.74   & 5.641 & H \\

\end{tabular}
\raggedright

Notes to Table 1: \\
Ca;  20 A/mm at H$\gamma$\\
Cb;  29 A/mm at H$\gamma$\\
Cc;  49 A/mm at H$\gamma$\\
Coud\'{e} (D) Series\\
DY; 13.6 A/mm (1st order); 6.7 A/mm (2nd order)\\
DZ; 31 A/mm (1st order); 15.6 A/mm (2nd order)\\
\end{table}

Table 2 lists other post-1947 results on high/low states from the
literature. In Tables 1 and 2 the date is given, as well as the cycle and
phase of each spectrum ($\phi$), based on the relation;
 \begin{equation}
\phi = (JD - 2430661)/2020.
 \end{equation}
Where JD is the Julian date. This uses the period and date of supposed
periastron passage ($\approx$ time of conjunction) in the orbit of Damineli
et al. (2000).  
We test this period below. In view of the work of Davidson et al. (2000)
which was mentioned in the Introduction, one is not constrained
in adopting any specific interpretation of the adopted zero-phase.
At least initially, it can be considered as simply a convenient
reference point.
In a few cases Table 2 shows that only the year and month of
the observation were published. The phase given then corresponds to the
middle of the month. The phase is thus uncertain in these cases by
$\sim$0.007. In no case is this of significance in the discussion.

\begin{table}
\centering
\caption{Spectral State (Misc)}
\begin{tabular}{rrrl}
\multicolumn{1}{c}{Date} & {Phase} & {State} & {ref}\\
 & & & \\
27.6.47  &  0.843 & H & 1,2 \\
19.4.48  &  0.990 & L & 1,2 \\
16.7.49  &  1.215 & H & 1,2 \\
30.4.61  &  3.346 & H & 3   \\
1.5.61   &  3.347 & H & 3   \\
2.5.61   &  3.347 & H & 3   \\
-.3.64   &  3.866 & H & 4   \\
-.2.65   &  4.032 & L & 4   \\
-.2.66   &  4.211 & H & 4  \\
10.3.74  &  5.672 & H & 5 \\
21.5.81  &  6.973 & L & 6 \\
-.6.81   &  6.986 & L & 7 \\
31.10.81 &  7.054 & L & 8 \\
1.11.81  &  7.054 & L & 8 \\
2.11.81  &  7.055 & L & 8 \\   
24.12.81 &  7.080 & L & 5 \\
-.12.82  &  7.256 & H & 7 \\
21.2.83  &  7.290 & H & 5 \\
4.4.85   &  7.673 & H & 6 \\
22.3.86  &  7.847 & H & 9 \\
-.1.87   &  7.996 & L & 7 \\
15.1.87  &  8.024 & L & 10 \\
20.6.92  &  8.977 & L & 11 \\
18.11.97 &  9.956 & H & 12 \\
24.11.97 &  9.958 & H & 13 \\
18.12.97 &  9.970 & L & 13 \\
25.12.97 &  9.974 & L & 12 \\
6.3.98   & 10.008 & L & 13 \\
23.3.98  & 10.018 & L & 12 \\
3.7.98   & 10.067 & H & 13 \\

\end{tabular}
\raggedright

References:\\
1 Gaviola (1953)\\ 
2 Viotti (1968)\\ 
3 Aller \& Dunham (1966)\\ 
4 Rodgers \& Searle (1967)\\ 
5 Zanella, Wolf \& Stahl (1984)\\ 
6 Bidelman, Galen \& Wallerstein (1993)\\ 
7 Damineli et al. (1994)\\
8 Paper~I\\ 
9 Hillier \& Allen (1992)\\ 
10 Altamore, Maillard \& Viotti (1994)\\ 
11 Damineli et al. (1998)\\ 
12 McGregor, Rathborne \& Humphreys (1999)\\ 
13 Wolf et al. (1999)\\ 

\end{table} 

The data from Tables 1 and 2 are shown plotted in Figs.~1 and 2.  Fig.~1
shows all the data whilst Fig.~2 shows the data for phases 0.8 to 0.2 on a
larger scale. The numbers are modified cycle numbers to avoid changing the
cycle at phase zero, in the middle of the low state. Thus, for example,
spectra with phases between 3.5 and 4.5 in the tables are plotted as cycle
number 4. The plot is clearly consistent with the recurrence of the low
excitation state in a periodicity close to that proposed by Damineli et al.
(2000) (which is, of course, partly based on some of these data). In
particular the three previously unreported low states in Thackeray's data
(1953-4, 1959, 1970) which are plotted as cycles 2, 3 and 5, fall in the
same phase range as the previously known low states. No low states have been
found between phases 0.1 and 0.9. 

Detailed examination of Fig.~2 shows that the phenomena do not repeat
exactly. There is a high state (cycle 10) at phase .067 whilst there are low
states at phase .080 (cycle 7) and .086 (cycle 2). If one adjusted the
period to remove this apparent anomaly then the high excitation point in
cycle 10 at phase .958 would be moved to a phase greater than the low
excitation point in cycle 7 currently shown at phase .977. Some caution is
necessary in this discussion since the criteria used (or their application)
may not have been exactly the same for all the spectra, especially for the
results of different authors which are taken from the literature.  Note that
the intermediate point in Fig.~2 at phase .929 (cycle 2) is the only
spectrum in Table 1 which suggests an intermediate classification. This
depends on a 4471He\,{\sc i}/4475[Fe\,{\sc ii}] estimate of low weight.
$\lambda$4471 is present (in a low state it is absent on these spectra) and
it is quite possible that the spectrum should actually be classified as in
the high state.  Leaving aside this intermediate point, the low state points
in cycle 7 show that some, and perhaps all, low states extend over a phase
range of at least 0.1. However, the high excitation phase points in cycle
10, if typical, show that the low state phase range cannot be significantly
greater than this.  Bearing in mind some possible variation from cycle to
cycle, we may conclude from Fig.~2 that the low state extends from phase
$\sim$ 0.970 to phase $\sim$ 0.085 and is centred at phase $\sim$ 0.025
which is about 50 days after periastron in the Damineli et al. (2000)
supposed orbit. The results of this section are further discussed in section
6.

\begin{figure*}
\centering
\epsfxsize=17.0cm
\epsffile{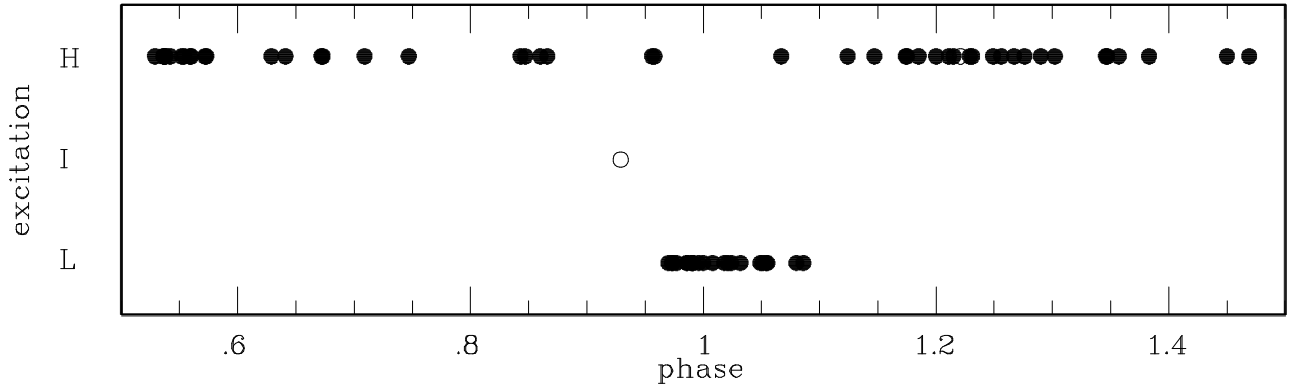}
\caption{The data from Tables 1 and 2 showing the high (H), Intermediate (I)
and low (L) states of ionization plotted against phase in the 2020 day
cycle. The open circles represent the two less certain assignments.}
\end{figure*}

\begin{figure*}
\centering
\epsfxsize=17.0cm
\epsffile{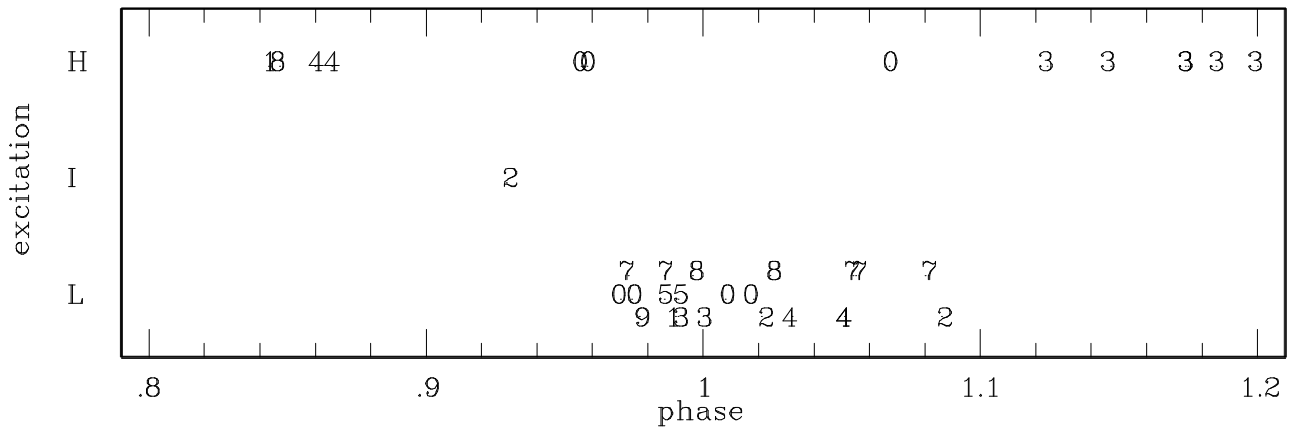}
\caption{As for Fig.~1, but for the restricted phase range, 0.8 to 0.2.
The cycle numbers defined in the text are shown (cycle 10 shown as 0).
The L state positions are spread vertically for clarity. }
\end{figure*}

\begin{figure*}
\centering
\epsfxsize=17.0cm
\epsffile{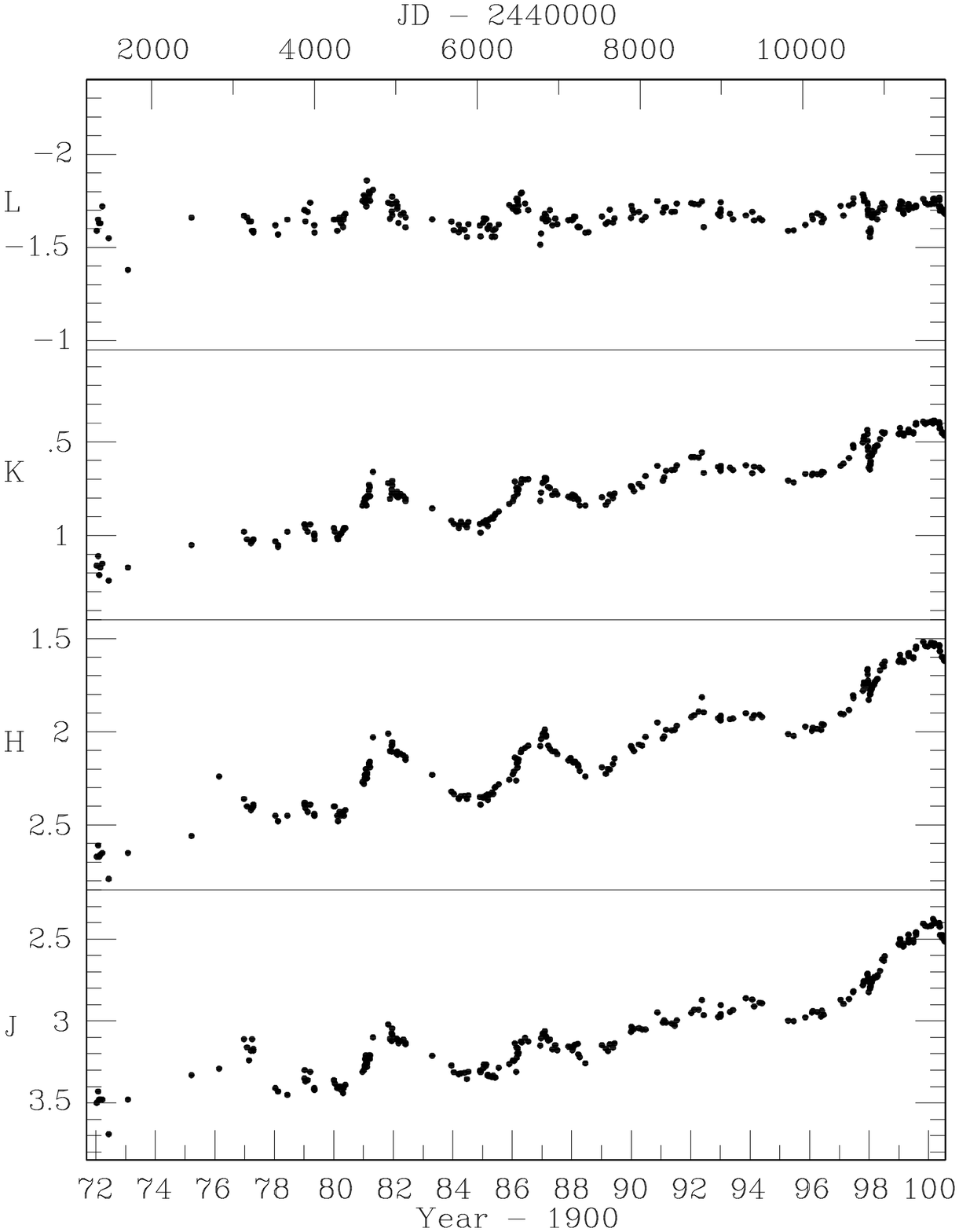}
\caption{Infrared light curves in $JHKL$ for 1972-2000.}
\end{figure*}
 
\section {Early Slit Spectra}
Walborn \& Liller (1977) have discussed early (19th century) objective prism
spectra of $\eta$ Car. Particularly important is the 1893 spectrum which
shows an F-type absorption spectrum. The first known slit spectrum was
obtained with the McClean (Victoria) telescope at the Cape in 1899 and was
described by Gill (1901). This is listed in Table 3 together with two plates
(also with the McClean telescope) of 1919 discussed by Lunt (1919), plates
of 1912-1914 from the Lick Southern station (Moore \& Sanford 1913) and
three Cape (McClean) plates of 1923.  The 1899 plate could not be located
but is reproduced in Gill's paper. The two 1919 plates are in the SAAO
archives and were examined (one is reproduced in Lunt's paper). The 1923
plates are recorded in the McClean plate book but are missing from the
archives. They do not seem to have been described in print. The SAAO
archives also have glass copies of the 1912-14 Lick plates including one
well exposed plate not discussed by the Lick workers. These glass copies
were apparently sent by G. Herbig to A.D. Thackeray.

The Cape 1919 plates and three of the Lick plates are of good quality, as
can be seen in the reproductions of some of them in the references cited
above. The spectra are striking in that although they are generally similar
to the more recent ones described in section 2, they all have 4471He\,{\sc i}
missing and can clearly be classified as in a low excitation state. In the
1919 plates it is possible that 4658[Fe\,{\sc iii}] is weakly present, but this is
not certain. Gill's (1901) line list of the 1899 plate at first looks
surprising since he measured a line at 4471.3 A. However, this spectrum (a 6
hour 10 min exposure over parts of 4 nights) was obtained during the test
phase of a new spectrograph and suffered from thermal instability problems.
[Fe\,{\sc ii}] and Fe\,{\sc ii} lines in the region of 4471 show that Gill's wavelengths are
$\sim$ 2A too small so that the true wavelength of the line is $\sim$4473.3.
Aller \& Dunham (1966) measured emission lines in a 1961 spectrum at 4472.9
Fe\,{\sc ii} (Intensity 4) and 4474.9 [Fe\,{\sc ii}] (Intensity 9). It is
likely that Gill's `4471' is a blend of these lines, especially as the ratio
Fe\,{\sc ii}/[Fe\,{\sc ii}] was probably greater in Gill's time than in 1961
(see section 4).

Table 3 shows the phases of the early plates using the Damineli et al.
(2000) elements (equation 1). It is clear that these are incompatible with
the restricted phase range of low excitation states shown in Fig.~2. It would
require a large period change (to $\sim$ 2150 days) to change the phase of
the 1913-19 observations to near zero and this would be inconsistent with
the discussion in sections 2 and 5 and with the 1899 spectrum.

\begin{table}
\centering
\caption{Early Slit Spectra}
\begin{tabular}{rrcr}
\multicolumn{1}{c}{Date} & {Phase} & {4471} & {Ref} \\ 
 & & & \\ 
14-17.4.1899 & 0.129 & (no) & 1\\
29.3.1912    & 0.471 &  -   & 2\\
11.3.1913    & 0.642 &  no  & 2\\
28.3.1913    & 0.651 &  no  & 2\\
2.3.1914     & 0.818 &  -   & 2\\
2.3.1914     & 0.818 &  -   & 2\\
11.3.1914      & 0.823 &  no  & 3\\
20-23.2.1919 & 0.718 &  no  & 4 \\
14-17.4.1919 & 0.745 &  no  & 4 \\
 & & & \\ 
15.6.1923    & 0.498 &  -   & 5 \\
18.6.1923    & 0.499 &  -   & 5 \\
19.6.1923    & 0.500 &  -   & 5 \\

\end{tabular}

\raggedright

References:\\
1 Gill 1901 \\
2 Moore \& Sanford 1914\\
3 Lick (unpublished)\\ 
4 Lunt 1919\\ 
5 Cape (missing plates)\\
\end{table}
We can conclude that whilst the low excitation states have occurred
regularly every $\sim$2020 days for the last 50 years at least, the 
situation was quite different 30 years before that. It seems
reasonable to assume that at these earlier times the spectrum of the
central object as viewed from the ground was always in the low state.
This result, together with the secular trends noted in the next
section must obviously be taken into account in any complete model
of the system.

Note that Gill (1901) also describes an objective prism plate of 1899
January, and Spencer Jones (1931) one of 1919 April. The SAAO archives
contain objective prism plates taken in May 1901. If slit spectra
taken in the period 1919-1947 exist they would be very valuable. We
are not, however, aware of any. 

\section{Long Term Spectral Changes}
 Baxandall (1919) and Lunt (1919) noted differences between the Cape
spectrum of 1899 and those of 1912-14 (Lick) and 1919 (Cape). Using modern
identifications these indicate that the Fe\,{\sc ii}/[Fe\,{\sc ii}] and
Ti\,{\sc II}/Fe\,{\sc ii} emission line ratios were greater in 1899 than
later. Thackeray (1953) noted that Ti\,{\sc ii} emission was weaker in
1951-53 than in 1919. he also noted (Thackeray 1967) that in 1961-5 Ti\,{\sc
ii} absorption was stronger than in Gaviola's 1949 observations.

It was shown in Paper~I that 2.06$\mu$m He\,{\sc i} emission was stronger in
1983, at phases 0.924, 0.950, and 0.082, than in the 1968 observation of
Westphal \& Neugebauer (1969) at phase 0.646.  Since Damineli (1996) shows
that the equivalent width of 10830 He\,{\sc i} emission is a minimum near
phase zero, one would have expected the opposite unless the strength of
He\,{\sc i} were increasing with time.  This together with the results of
the last section suggest a secular increase in the strengths of high
excitation lines (He\,{\sc i} etc.). Other evidence for this is briefly
summarized by Damineli et al. (1999).
 
In contrast, Paper~I also showed that the equivalent width of Br
$\gamma$ was about the same in 1981 as 1968. Damineli et al.
(1997) found that Pa $\gamma$ varied little in strength with respect to the
continuum in the 2020 day cycle. These results suggest that the hydrogen
emission varies with the continuum both in the cyclical variations and with
a secular trend. Note, however, that Aitken et al. (1977) who observed at
phase 0.08 in 1976 found B$\gamma$ to be only half the strength of that
found in Paper~I at about the same phase in 1981. This might be due to the
fact that Aitken et al. used a much smaller aperture (3.4 $\times$ 4.8
arcsec) than either Westphal \& Neugebauer or Paper~I; although in Paper~I no
difference was found between observations though apertures of 12 and 36
arcsec.
\begin{table}
\centering
\caption{Near-Infrared Photometry}
\begin{tabular}{rrrrr}
\multicolumn{1}{c}{JD--2440000}&\multicolumn{1}{c}{$J$}&
\multicolumn{1}{c}{$H$}&\multicolumn{1}{c}{$K$}&\multicolumn{1}{c}{$L$}\\
\multicolumn{1}{c}{(day)}&\multicolumn{4}{c}{(mag)}\\
 9379.52 & 2.869 & 1.927 &  0.667 &--1.691  \\
 9402.54 & 2.910 & 1.911 &  0.632 &--1.645  \\
 9468.41 & 2.887 & 1.909 &  0.638 &--1.656  \\
 9498.30 & 2.890 & 1.920 &  0.650 &--1.647  \\
 9820.32 & 2.997 & 2.012 &  0.706 &--1.590  \\
 9887.31 & 3.001 & 2.023 &  0.716 &--1.592  \\
10031.58 & 2.977 & 1.972 &  0.671 &--1.621  \\
10111.58 & 2.948 & 1.995 &  0.677 &--1.672  \\
10123.53 & 2.938 & 1.977 &  0.669 &--1.651  \\
10178.39 & 2.944 & 1.984 &  0.672 &--1.684  \\
\\[-1mm]
10224.42 & 2.973 & 1.991 &  0.675 &--1.672  \\
10234.22 & 2.937 & 1.959 &  0.660 &--1.635  \\
10256.26 & 2.962 & 1.962 &  0.662 &--1.655  \\
10464.57 & 2.870 & 1.903 &  0.629 &--1.723  \\
10499.49 & 2.895 & 1.905 &  0.616 &--1.671  \\
10567.36 & 2.865 & 1.883 &  0.586 &--1.725  \\
10618.19 & 2.823 & 1.804 &  0.518 &--1.736  \\
10623.30 & 2.817 & 1.819 &  0.533 &--1.764  \\
10738.62 & 2.781 & 1.780 &  0.504 &--1.786  \\
10746.62 & 2.758 & 1.750 &  0.484 &--1.774  \\
\\[-1mm]
10750.62 & 2.756 & 1.734 &  0.472 &--1.749  \\
10755.62 & 2.760 & 1.741 &  0.470 &--1.761  \\
10792.56 & 2.714 & 1.671 &  0.437 &--1.740  \\
10796.50 & 2.710 & 1.663 &  0.458 &--1.699  \\
10798.44 & 2.719 & 1.693 &  0.495 &--1.728 \\
10800.49 &       & 1.738 &  0.528 &          \\
10801.53 & 2.731 & 1.725 &  0.528 &--1.702 \\
10802.61 & 2.742 & 1.735 &  0.547 &--1.672 \\
10805.61 & 2.758 & 1.766 &  0.578 &--1.682 \\
10809.60 & 2.824 & 1.829 &  0.638 &--1.585 \\
\\[-1mm]
10824.57 & 2.802 & 1.801 &  0.648 &--1.556 \\
10827.56 & 2.794 & 1.798 &  0.620 &        \\
10830.56 & 2.765 & 1.780 &  0.613 &--1.602 \\
10831.52 & 2.789 & 1.783 &  0.622 &--1.580 \\
10833.59 & 2.782 & 1.780 &  0.603 &--1.590 \\
10848.55 & 2.749 & 1.760 &  0.567 &--1.697 \\
10849.63 & 2.762 & 1.768 &  0.572 &--1.660 \\
10850.65 & 2.753 & 1.754 &  0.566 &--1.665 \\
10852.56 & 2.753 & 1.762 &  0.563 &--1.666 \\
10877.61 & 2.738 & 1.747 &  0.552 &--1.678 \\
\\[-1mm]
10879.48 & 2.730 & 1.738 &  0.538 &--1.681 \\
10887.47 & 2.734 & 1.731 &  0.526 &         \\
10912.38 & 2.730 & 1.718 &  0.518 &--1.651 \\
10918.36 & 2.720 & 1.716 &  0.520 &--1.688 \\
10948.30 & 2.693 & 1.670 &  0.485 &--1.710 \\
10976.22 & 2.624 & 1.639 &  0.448 &--1.737 \\
10996.20 & 2.633 & 1.650 &  0.457 &--1.702 \\
11006.28 & 2.603 & 1.623 &  0.450 &--1.720 \\
11179.61 & 2.528 & 1.623 &  0.458 &--1.707 \\
11186.56 & 2.534 & 1.616 &  0.456 &--1.709 \\
\\[-1mm]
11194.62 & 2.498 & 1.586 &  0.426 &--1.729 \\
11200.62 & 2.503 & 1.609 &  0.452 &--1.748 \\
11221.63 & 2.535 & 1.623 &  0.458 &--1.742 \\
11233.48 & 2.545 & 1.618 &  0.457 &--1.687 \\
11238.41 & 2.528 & 1.625 &  0.461 &--1.714 \\
11239.48 & 2.540 & 1.627 &  0.467 &--1.680 \\
11298.25 & 2.472 & 1.581 &  0.441 &--1.734 \\
11299.49 & 2.500 & 1.583 &  0.434 &--1.711 \\
11300.30 & 2.510 & 1.595 &  0.450 &--1.704 \\
11302.26 & 2.520 & 1.576 &  0.449 &--1.723 \\
\multicolumn{5}{l}{continued ...}
\end{tabular}
\end{table}
\setcounter{table}{3}
\begin{table}
\centering
\caption{continued}
\begin{tabular}{rrrrr}
\multicolumn{1}{c}{JD--2440000}&\multicolumn{1}{c}{$J$}&
\multicolumn{1}{c}{$H$}&\multicolumn{1}{c}{$K$}&\multicolumn{1}{c}{$L$}\\
\multicolumn{1}{c}{(day)}&\multicolumn{4}{c}{(mag)}\\
11355.28 & 2.504 & 1.599 &  0.449 &--1.714 \\
11358.21 & 2.519 & 1.607 &  0.456 &--1.715 \\
11389.19 & 2.461 & 1.555 &  0.408 &--1.725 \\
11391.19 & 2.473 & 1.544 &  0.408 &--1.718 \\
11392.19 & 2.457 & 1.543 &  0.399 &--1.721 \\
11479.60 & 2.405 & 1.517 &  0.393 &--1.761 \\
11501.60 &       &       &  0.403 &       \\
11505.61 & 2.419 & 1.540 &  0.405 &--1.739 \\
11536.60 & 2.422 & 1.542 &  0.397 &--1.730 \\
11572.57 & 2.418 & 1.530 &  0.396 &--1.732 \\
\\[-1mm]
11575.55 & 2.419 & 1.521 &  0.393 &--1.734 \\
11599.47 & 2.376 & 1.536 &  0.402 &--1.760 \\
11613.37 & 2.392 & 1.539 &  0.386 &--1.762 \\
11615.34 & 2.392 & 1.525 &  0.394 &--1.754 \\
11617.41 & 2.403 & 1.527 &  0.396 &--1.735 \\
11674.27 & 2.411 & 1.535 &  0.394 &--1.718 \\
11676.27 & 2.401 & 1.535 &  0.407 &--1.768 \\
11678.29 & 2.413 & 1.541 &  0.405 &--1.754 \\
11683.34 & 2.424 & 1.567 &  0.430 &--1.723 \\
11685.39 & 2.474 & 1.567 &  0.424 &--1.697 \\
\end{tabular}
\end{table}

\section{Infrared Photometry}
 Monitoring of $\eta$ Car in $JHKL$ at SAAO in the period 1972-94
established a quasi-cyclical variation with a period of the order of 5 years
superposed on a secular increase in brightness (see Paper~II). Table 4
contains the results of a continuation of this series of observations for
the period 1994-2000. As in all recent work in this series, the observations
were made with the SAAO MkII photometer on the 0.75m reflector at SAAO,
Sutherland and using a 36\,arcsec diaphragm. The individual results are
accurate to better than $\pm$0.03 in $JHK$ and $\pm$ 0.05 in $L$. The
results are in the system defined by the standard stars of Carter (1990).

Fig.~3 shows the photometry for the period 1972-2000. The secular increase
in brightness has continued and its rate has increased (see Davidson et al.
1999). This long term brightening is most marked at $J$. The maximum in the
2020 day cycle expected near 1998.0 is not seen in $JHK$. In fact, the light
curves peak only in early 2000 where this data set ends. This can be
plausibly attributed to the increased rate of secular brightening. Perhaps
the most striking feature of the light curves is the marked dip near 1998.0
which was originally noted by Whitelock \& Laney (1999). This is illustrated
in more detail in Fig.~4.  The minimum of this dip occurs between
JD\,2450809 and 2450824 and is evidently within a few days of JD\,2450815.
This is at phase 0.977 or 46 days before zero-phase on the Damineli et al.
elements (equation 1).  Fig.~5 compares the region near zero-phase in this
cycle (cycle 10) at $K$ with the observations in the two previous cycles. 
Unfortunately the observations are sparse at these earlier epochs.
Nevertheless, there is a drop in brightness consistent with the beginnings
of a dip in the cycle 9 data and of an emergence from a dip in the cycle 8
data. In cycle 7 there is a gap of more than 180 days in the data at the
crucial phase and before that the data are too sparse to use in identifying
dips. The predicted minimum in cycle 8 is at JD\,2446775 which is in fact
the date of the faintest point in this cycle. Thus if the dip occurs with
strict regularity the period cannot be less than 2020 days. The predicted
minimum in cycle 9 is at JD\,2448795, whilst the faintest point is at
JD\,2448783. Thus the period cannot be made longer by more than 12 days at
the most.  It may therefore be concluded that although the data are rather
sparse except in cycle 10, the evidence from cycles 8 and 9 indicates that
the dip does occur regularly with a period close to 2020 day. Note that
although the above discussion refers to the $K$ magnitudes the results are
the same in the other three colours.

Since the $L$ magnitude is less affected by the secular trend than $JHK$,
one can more safely look at the 2020 day variations at this wavelength.
Fig.~6 shows the $L$ data folded on this period with a linear trend
removed and with a third order sine curve fitted.  The scatter ($\sigma$ =
0.042 mag.) apparent in this plot is partly real and partly observational. 
The fitted curve is relatively flat except for a `hump' between phases
$\sim$ 0.7 and 0.2, which also contains the dip.
 
The dip shown by the infrared observations (Fig.~4) is at about the
same phase as that shown by the X-ray observations 
(Ishibashi et al. 1999). The principal difference being that the 
X-ray dip is much deeper. The dip in J begins about 35 days (0.017 of
a cycle) later than that in the X-rays and coincides with the phase at which
the  X-rays reach minimum. Note that the epoch of minimum increases
slightly with increasing wavelength through the infrared as can be
seen from Fig.~4. This together with the X-ray results suggests a possible
general dependence of the phase of the dip on wavelength. 
However, the effect in the infrared may be due, at least partially, to
the steeper secular increase in brightness at the shorter wavelengths.
There is also
a small dip in the visual light curve (van Genderen et al. 1999), though
the relevant epoch is not well enough sampled to define its depth and shape. 

\begin{figure}
\centering
\epsfxsize=8.4cm
\epsffile{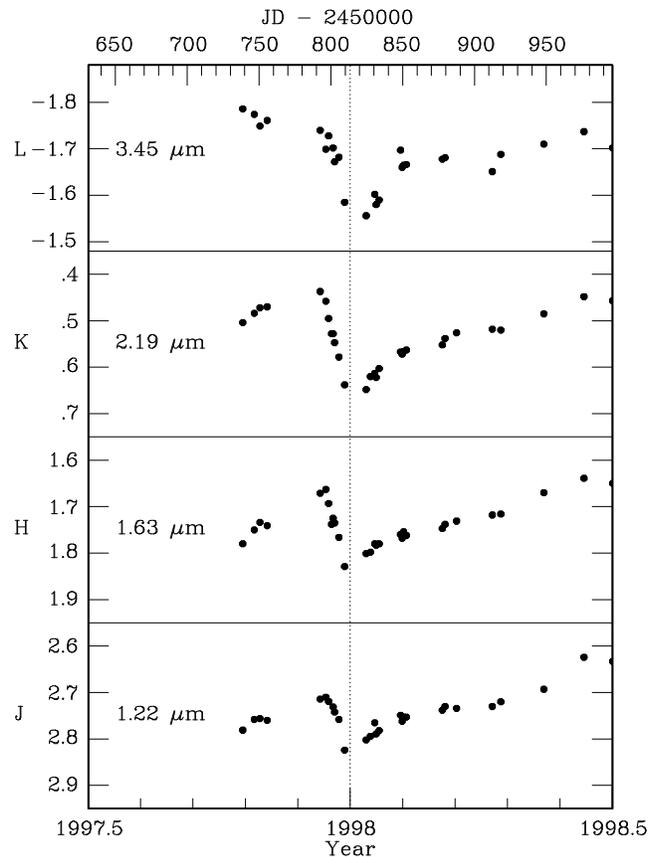}
\caption{Infrared light curves in $JKHL$ near the time of the 1998 dip.
The minimum in the dip (phase 0.977) is shown by the vertical line.}
\end{figure}

\begin{figure}
\centering
\epsfxsize=8.4cm
\epsffile{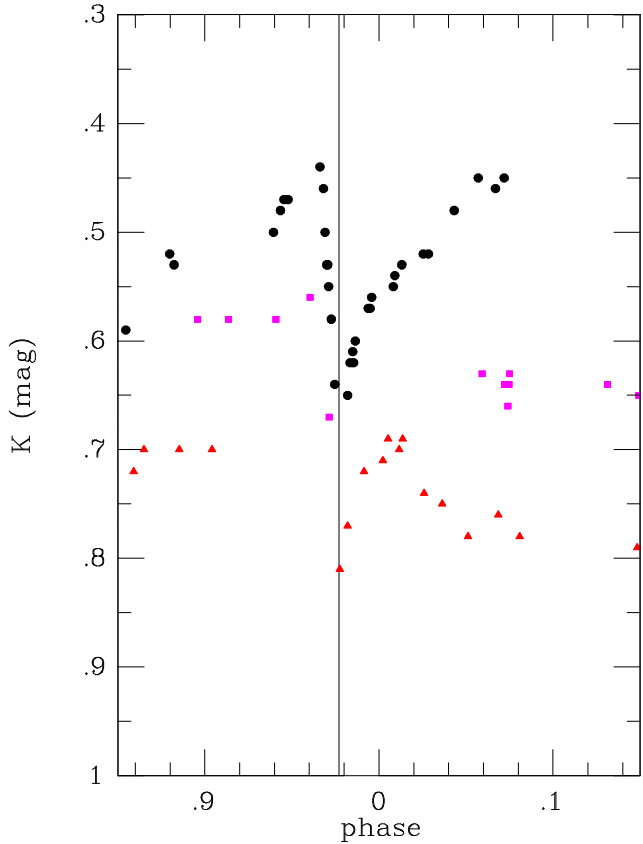}
\caption{The light curve at $K$ near phase 0.0 at several epochs,
triangles - cycle 8; squares - cycle 9; filled circles - cycle 10.
The vertical line shows the time of minimum (phase 0.977).}
\end{figure}

Fig.~7 shows the $L$ light curve for phases 0.8 to 0.2 with the linear
trend and the third order sine curve shown in Fig.~6 removed. The dip is seen
to extend from phase $\sim$ 0.96 to $\sim$ 0.06. The centre of the dip is at
phase $\sim$ 0.01 and the deepest point at phase 0.977 (the point shown at
this phase in Fig.~7 is from cycle 8 at JD\,2446775). The asymmetric shape of
the dip (steep drop, slower rise) is qualitatively similar to that shown by
the X-rays (Ishibashi et al. 1999). 

\section{Discussion}
\subsection{Periodic Variations}
 The interpretation of data on $\eta$ Car is complicated by the fact that
different types of observation often refer to physically separate regions of
the object. Optical aperture photometry refers mainly to the integrated
light from the homunculus.  This light is believed to be dust-scattered
radiation from the central object. The mid-infrared ($\sim$ 10$\mu$m)
radiation is mainly dust emission from the homunculus. Near infrared
($\sim$2$\mu$m) radiation comes from a region much more concentrated to the
central object than the optical light. The rich, sharp, emission line
spectrum which is so prominent on ground-based slit spectra of the central
region and shows the high/low excitation states discussed above, has been
found from HST observations (Davidson et al. 1997) to come, not from the
central object itself, but from nebular knots (their B, C, D) about 0.2
arcsec away.

The new epochs of low excitation reported in section 2 together with those
already known are clearly consistent with the periodicity of 2020 days
suggested by Damineli et al., at least over the last 50 years. As already
noted, Damineli et al. (1997) found radial velocity changes in the
Pa$\gamma$ line which they suggested were due to motion of a star (the
primary) in a highly eccentric binary orbit.  In this orbit periastron is
within 0.001 of inferior conjunction.  Davidson et al. (2000) point out that
much of the emission observed in Pa$\gamma$ from the ground comes from
surrounding nebulosity, not the central object itself. Their HST
observations at two epochs do not support the Damineli et al. orbit.

Despite the results of Davidson et al. (2000), a binary model of some kind
is still probably the most satisfactory explanation of the 2020 day
periodicity. Such a conclusion is supported by the infrared and X-ray dips
which seem to point to some sort of eclipse phenomenon. Whitelock \& Laney
(1999) noted that that the infrared light curve bore a rather striking
qualitative resemblance to the optical light curve of some classical
cataclysmic variables where a dip is superposed on a `hump' in the light
curve (see section 5 and Fig.~6 of the present paper). In these systems much
of the light is due to a hot spot on a disc around a white dwarf; the spot
being heated by infalling matter from a secondary star. The hump is due to
the varying aspect of the spot as it moves in the orbit and the dip is due
to its eclipse by the secondary. The infrared (Fig.~7) and X-ray (Ishibashi
et al. 1999) light curves of $\eta$ Car even show evidence of a two step
structure similar to that seen in the eclipse light curves of some
cataclysmic variables (e.g. Z Cha, see Cook \& Warner 1984). In the case of
$\eta$ Car the `hot spot' could either be on a disc or else be a hot region
in an interacting wind model and the eclipse due either to an optically
thick plasma or a star. This type of model does not necessarily require an
eccentric orbit. It remains to be determined how much the `hot spot'
contributes to the total luminosity of the system. The depth of the eclipse
shows that at least 20 percent of the near-infrared flux and essentially all
of the X-ray flux is generated in the `hot spot'.

Davidson (1997) has suggested that the modulation of the near infrared flux
in the 2020 day cycle is due to variation in free-free emission.  This is
evidently consistent with a `hot spot' model. Smith \& Gehrz (2000) find,
from high resolution images, that the flux near 2$\mu$m was more centrally
concentrated when the object was brighter at this wavelength, though it is not
entirely clear how much this is related to the 2020 day periodicity and how
much to the secular variations.  The sharp emission-line spectrum, coming
from a region separated from the central object, is presumably at least
partially excited by radiation from the `hot spot'. This spectrum will thus
vary, in the 2020 day cycle, according to the relative positions of the spot
(and/or its intensity) and the excited region. Hence the orbit must be
eccentric, unless the ejecta responsible for the sharp line spectra are
distributed asymmetrically about the central object.  Davidson et al. (1997)
show that the ejecta (C,D) are moving towards us relative to the central
object indicates that this latter condition is fulfilled.
\begin{figure*}
\centering
\epsfxsize=15.0cm
\epsffile{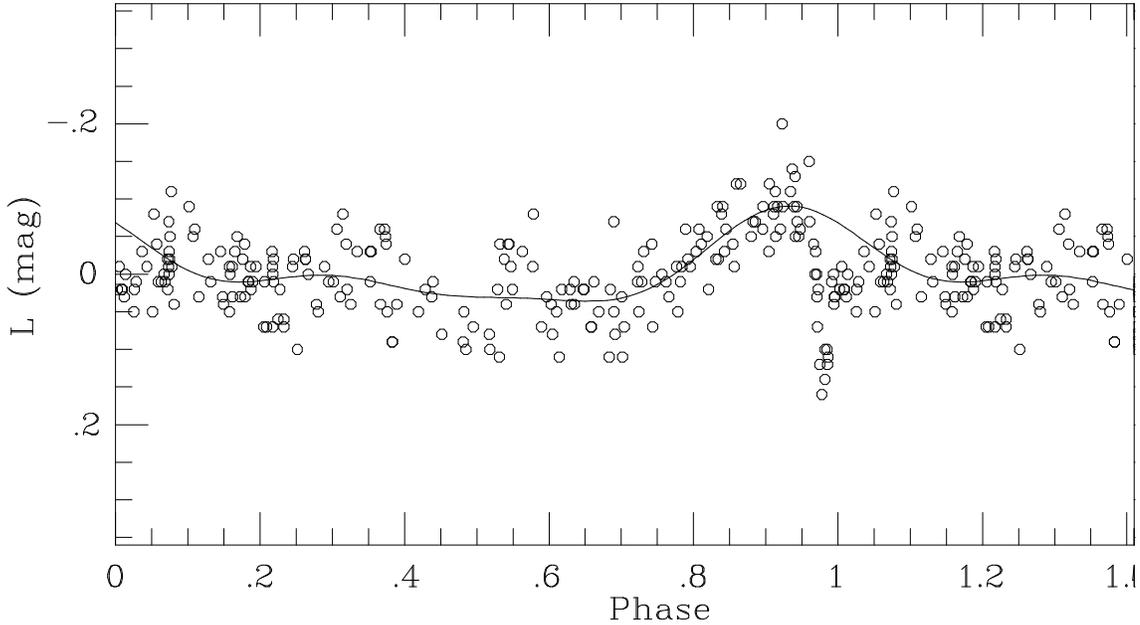}
\caption{The phased light curve at $L$, assuming a period of 2020 day
and with a linear trend, of 0.002 mag$\rm \, yr^{-1}$, removed.
A third order sine fit is shown ($\sigma=0.042$\,mag).}
\end{figure*} 

It would appear to be significant that the phase and duration of the eclipse
is the same as that of the low excitation state. Any phase offset between
the two is quite small. The X-ray flux which is being eclipsed is presumably
generated in the `hot spot', the sharp emission lines come from a region
separated from this. The infrared flux may come from either or both these
regions.  The similar lengths of the eclipse and of the low excitation state
imply that the emission-line region can only subtend a small angle as seen
from the central source. Furthermore, the fact that these two phenomena
occur at the same phase indicates that the direction from the central source
to the emission-line region(s) can only make a small angle with the line of
sight from the central source to the observer.
Davidson et al. (1997) show that the emission line regions C and D lie in
the equatorial plane of the bipolar system. This is evidently consistent
with our eclipse model since we would expect the orbital plane of the binary
to coincide with the plane of symmetry of the bipolar lobes. The inclination
of the equatorial plane to the plane of the sky is $i = 57^o \pm
10^o$ (Davidson \& Humphreys 1997). Stevens \& Pittard (1999) obtained
significant X-ray eclipses in a model which adopted
$i = 60^o$ consistent with this.
However the referee (Professor Davidson) informs us that
Davidson et al. (to be published) have used HST/STIS data to revise
the inclination from $57^{o}$ to $40^{o}$ which makes it more
difficult to maintain an eclipse model with the orbit in this plane.

We have stressed the hot-spot model because the observed light curve
is strikingly similar to that of a cataclysmic variable, though on
a vastly different scale. Evidently this model must remain speculative
pending a quantitative approach. The referee has in fact suggested
that one should consider an alternative model (Davidson 1999) in which
the spectroscopic low states etc. are the result of a secondary
star entering a circumstellar disc round the primary or else inducing
sudden mass ejection from the primary.

\begin{figure}
\centering
\epsfxsize=8.4cm
\epsffile{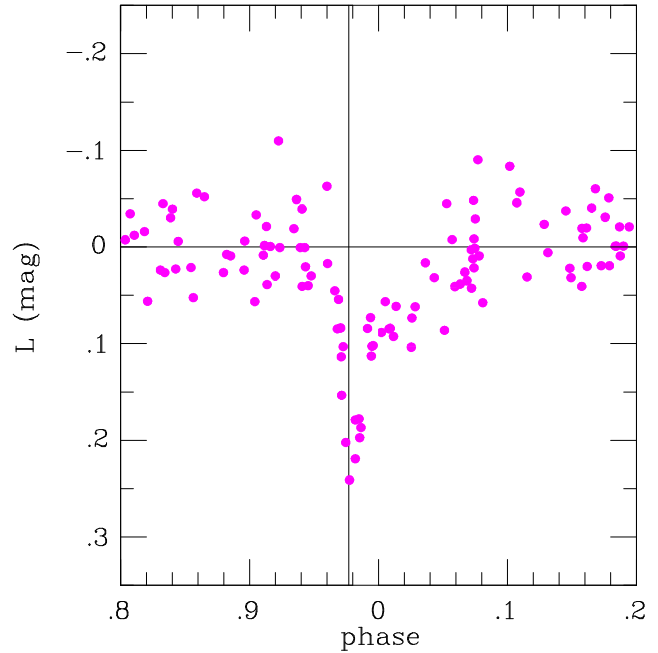}
\caption{The $L$ light curve from phase 0.8 to 0.2 with a linear trend 
and the third order sine curve of Fig.~6 removed.}
\end{figure}

\subsection{Secular Variations}
 The gradual brightening of the $\eta$ Car system over much of the
20th century has generally been attributed to a gradual decrease in
reddening as the dust formed in the great eruption dissipates. Whilst this
seems plausible as a general explanation, the details are undoubtedly
complex. Over the last 30 years, at least, the object has brightened by
about the same amount in $JHK$ as in the visual region and this has
suggested a large-particle model (neutral extinction). However, the visual
brightness, as normally estimated, is dominated by scattered light from the
homunculus whereas the $JHK$ emission is more concentrated to the central
object.  It is not entirely clear how the visual brightness of the
homunculus will change as it expands and both absorption and scattering
decrease. The recent increase in the rate of brightening (Davidson et al.
1999) is in any case incompatible with a simple model of decreasing
absorption due to the steady expansion of a dust shell. Furthermore it is
striking that this increase is similar in the visual and near infrared. It
remains possible that some of the observed variations are due to changes in
the central engine itself, as suggested in Paper~I, and this would seem
possible on a `hot spot' model.

That the brightening is not simply due to the clearing of obscuration is
also evident from the fact that there are long term physical changes in the
system. Not only is the homunculus expanding, but the ejecta which
contribute the sharp-line spectrum are moving away from the central object
(Davidson et al. 1997) and are undergoing long term changes which were
reviewed in section 2. The secular increase in the strength of He\,{\sc i}
reflects an increase in excitation. The evidence also suggests an increase
in [Fe\,{\sc ii}]/Fe\,{\sc ii} between 1899 and 1913 which presumably
indicates a decrease in density in the emitting region. This ratio does not
seem to have changed significantly since 1913.

\section{conclusions}
 The photometric and spectroscopic results discussed in this paper seem most
naturally explicable by an interacting-binary model of some kind. A binary
model involving a `hot spot' is basically that proposed originally by Bath
(1979); though the long period presumably indicates that in the case of
$\eta$ Car, the interaction involves a stellar wind from at least one of the
stars rather than Roche-lobe overflow.
It is evident that a full quantitative model of the $\eta$ Car system
is still a matter for the future. However the results of the present
paper provide considerable constraints for any such model.

\subsection*{Acknowledgements}
 We are grateful to Dave Laney, Tom Lloyd Evans, Enrico Olivier and Ronald
Heijmans for making some of the $JHKL$ observations.  The re-examination of
the archival spectra arose out of discussions with Roberta Humphreys. We
thank the referee (Kris Davidson) for some pertinent comments. This paper is
based partly on observations made from the South African Astronomical
Observatory.


\begin{thebibliography}{} 
\bibitem[]{} Aitken D.K., Jones B., Bregman J.D., Lester D.F.,
Rank D.M., 1977, ApJ, 217, 103
\bibitem[]{} Aller L.H., Dunham T., 1966, ApJ, 146, 126
\bibitem[]{} Altamore A., Maillard J.-P., Viotti R., 1994,
A\&A, 292, 208
\bibitem[]{} Bath G.T., 1979, Nature, 282, 274
\bibitem[]{} Baxandall F.E., 1919, MNRAS, 79, 619
\bibitem[]{} Bidelman W.P., Galen T.A., Wallerstein G., 1993,
PASP, 105, 785
\bibitem[]{} Carter B.S., 1990, MNRAS, 242, 1
\bibitem[]{} Cook M.C., Warner B., 1984, MNRAS, 207, 705
\bibitem[]{} Damineli A., 1996, ApJ, 460, L49
\bibitem[]{} Damineli A., Viotti R., Cassatella A., Baratta G.B.,
1994, SSR, 66, 211
\bibitem[]{} Damineli A., Conti P.S., Lopes D.F., 1997, New Ast., 2, 107
\bibitem[]{} Damineli A., Stahl O., Kaufer A., Wolf B., Quast G.,
Lopes D.F., 1998, A\&AS, 133, 299
\bibitem[]{} Damineli A., Viotti R., Kaufer A., Stahl O.,
Wolf B., de Ara\'{u}jo F.X., 1999, in: 
Eta Carinae at the Millennium, 
J.A. Morse et al. (eds.), ASP Conf. Ser., 179, p.~196
\bibitem[]{} Damineli A., Kaufer A., Wolf B., Stahl O.,
Lopes D.F., de Ara\'{u}jo F.X., 2000, ApJ, 528, L101
\bibitem[]{} Davidson K., 1997, New Ast., 2, 387
\bibitem[]{} Davidson K., 1999, in: J.A. Morse et al. (eds.)
Eta Carinae at the Millennium, ASP Conf. Ser., 179, 304
\bibitem[]{} Davidson K., Humphreys R.M., 1997, ARAA, 35, 1
\bibitem[]{} Davidson K., et al., 1997, AJ, 113, 335
\bibitem[]{} Davidson K., et al., 1999, AJ, 118, 1777
\bibitem[]{} Davidson K., Ishibashi K., Gull T.R., Humphreys R.M.,
Smith N., 2000, ApJ, 530, L107
\bibitem[]{} Gaviola E., 1953, ApJ, 118, 234
\bibitem[]{} Gill D., 1901, Proc. Roy. Soc. (Lond) 68, 456
and 1901, MNRAS, 61, Appendix p.~66
\bibitem[]{} Hillier D.J., Allen D.A., 1992, A\&A, 262, 153
\bibitem[]{} Ishibashi K., Corcoran M.F., Davidson K., 
Swank J.H., Petre R., Drake S.A., Damineli A., White S., 1999,
ApJ, 524, 983   
\bibitem[]{} Lunt J., 1919, MNRAS, 79, 621
\bibitem[]{} McGregor P.J., Rathborne J.M., Humphreys R.M.,
in: Eta Carinae at the Millennium, J.A. Morse et al. (eds)
ASP Conf. Ser., 179, p.~236
\bibitem[]{} Moore J.H., Sanford R.F., 1913, Lick Obs. Bull. 8, 55 (No. 252)
\bibitem[]{} Morse J.A., Humphreys R.M., Damineli A., (eds.), 1999, 
Eta Carinae at the Millennium, ASP Conf. Ser., 179
\bibitem[]{} Rodgers A.W., Searle L., 1967, MNRAS, 135, 99
\bibitem[]{} Smith N., Gehrz R.D., 2000, ApJ, 529, L99
\bibitem[]{} Spencer Jones H., 1931, MNRAS, 91, 794
\bibitem[]{} Stevens I.R., Pittard J.M., 1999, in: Eta
Carinae at the Millennium, J.A. Morse et al. (eds.) 
ASP Conf. Ser., 179, p.~295 
\bibitem[]{} Thackeray A.D., 1953, MNRAS, 113, 211
\bibitem[]{} Thackeray A.D., 1967, MNRAS, 135, 51
\bibitem[]{} Thackeray A.D., 1977, Mem.RAS, 83,1
\bibitem[]{} van Genderen A.M., Sterken C., de Groot M., Burki G.,
 1999, A\&A, 343, 847
\bibitem[]{} Viotti R., 1968, Mem. Soc. Ast. It., 39, 105 
\bibitem[]{} Walborn N.R., Liller M.H., 1977, ApJ, 211, 181
\bibitem[]{} Westphal J.A., Neugebauer G., 1969, ApJ, 156, L45
\bibitem[]{} Whitelock P.A., Feast M.W., Carter B.S., Roberts G.,
Glass I.S., 1983, MNRAS, 203, 385, Paper~I 
\bibitem[]{} Whitelock P.A., Feast M.W., Koen C., Roberts G.,
Carter B.S., 1994, MNRAS, 270, 364, Paper~II 
\bibitem[]{} Whitelock P.A., Laney C.D., 1999, in:  J.A. Morse et al. (eds.)
Eta Carinae at the Millennium, ASP Conf. Ser., 179, p.~258
\bibitem[]{} Wolf B., Kaufer A., Stahl O., Damineli A., 1999,
in: Eta Carinae at the Millennium, J.A. Morse et al. (eds.),
ASP Conf. Ser., 179, p.~243
\bibitem[]{} Zanella R., Wolf B., Stahl O., 1984, A\&A, 137, 79
\end{thebibliography}
\end{document}